\pdfoutput=1
\documentclass{JINST}

\usepackage{overpic}
\usepackage{booktabs}
\usepackage{gensymb}

\title{The artificial retina processor for track reconstruction at the LHC crossing rate}

\author{
A.~Abba$^a$, 
F.~Bedeschi$^b$, 
M.~Citterio$^a$, 
F.~Caponio$^a$, 
A.~Cusimano$^a$,
A.~Geraci$^a$,
P.~Marino$^b$,
M.J.~Morello$^b$,
N.~Neri$^a$,
G.~Punzi$^b$,
A.~Piucci$^b$,
L.~Ristori$^{b,c}$,
F.~Spinella$^b$,
S.~Stracka$^b$,~and 
D.~Tonelli$^d$\thanks{Corresponding author.}\\
\llap{$^a$} Politecnico and INFN, Milano,\\ Via Celoria 16, 20133 Milano, Italy\\
\llap{$^b$} University, Scuola Normale Superiore, and INFN Pisa,\\ Largo Pontecorvo 3, 56127 Pisa, Italy\\
\llap{$^c$} Fermilab,\\ PO Box 500, 60510 Batavia, IL, U.S.A. \\
\llap{$^d$} CERN,\\ CH-1211 Geneva 23, Switzerland\\

E-mail: \email{diego.tonelli@cern.ch}}

\abstract{We present results of an R\&D study for a specialized processor capable of precisely reconstructing, in pixel detectors, hundreds of charged-particle tracks from high-energy collisions at 40 MHz rate. We apply a highly parallel pattern-recognition algorithm, inspired by studies of the processing of visual images by the brain as it happens in nature, and describe in detail an efficient hardware implementation in high-speed, high-bandwidth FPGA devices. This is the first detailed demonstration of reconstruction of offline-quality tracks at 40 MHz and makes the device suitable for processing Large Hadron Collider events at the full crossing frequency.}


\begin{document}


\section{Introduction}\label{sec:intro}

Charged-particle trajectories (tracks) are among the most physics-rich quantities typically available in collider experiments. Tracks encapsulate kinematic, lifetime, and charge  information in a handful of parameters, which are usually measured accurately, owing to the high precision of current position-sensitive detectors. Track information is therefore attractive to discriminate in real time the $10^{-3}-10^{-5}$ fraction of events that are typically stored for further processing in high-rate hadron collisions.  However, real-time track reconstruction at high event rates is a major challenge that implies doing pattern recognition in the presence of large combinatorics and handling a large information flow. This calls for highly parallel pattern-recognition algorithms that only use the subset of information needed to efficiently reconstruct tracks. 
Devices dedicated to online track-reconstruction in hadron collisions were employed since the early '80s~\cite{Famp}. In the '90s, the Collider Detector at Fermilab (CDF) used pattern-matching algorithms implemented into field-programmable gate-arrays (FPGA) to reconstruct two-dimensional tracks from clusters of aligned hits in a large
drift-chamber~\cite{XFT}. In 2001 the silicon vertex trigger~\cite{SVT} implemented fast and efficient pattern-matching using a custom-made processor, the associative memory, that connected the drift-chamber tracks with silicon-detector information and made available two-dimensional tracks with offline-like resolution at 30--100 Hz within the $20$ $\mu$s latency of the second level of CDF's three-stage trigger. Online track triggers, based on content-addressable memories, were also used in the less demanding environment of proton-electron collisions~\cite{H1}. Real-time track reconstruction would greatly benefit the experiments at the Large Hadron Collider (LHC). As of year 2020, higher LHC energies and luminosities will severely challenge the experiments' data acquisition and event reconstruction capabilities. In addition, the large number of interactions per bunch crossing, and the complexity of each event will reduce the discriminating power of usual experimental signatures, such as charged leptons with a large momentum transverse to the beam (transverse momentum, $p_T$) or significant unbalances in total event $p_T$.  We realize a detailed bit-wise simulation of an implementation into FPGAs of a novel neurobiology-inspired pattern-recognition algorithm, the \emph{artificial retina}, which proves particularly suited for real-time tracking in high-luminosity LHC conditions.

\section{Artificial retina tracking}\label{sec:retina}

The artificial retina tracking algorithm~\cite{retinaNIM} was inspired by the understanding of the mechanism of visual receptive fields in the mammals' eye~\cite{HubertWiesel}, whose functionalities have recently been shown to mirror those employed in high-speed digital data reduction~\cite{PunziDelViva}. Each neuron dedicated to vision is tuned to recognize a specific simple shape on a specific region of the mammals' retina, the \emph{receptive field}. The neuron response intensity to a visual stimulus is proportional to the degree of similarity between the shape of the stimulus and the shape for which the neuron is tuned to. Hence, each neuron reacts to the stimulus with different intensity. The brain extracts the first higher-resolution information on the basic geometric features of the stimulus by interpolation between the neuron responses, within a time of approximately 30 ms in humans. For a typical neuron firing frequency of 1 kHz, this corresponds to approximately 30 processing cycles. At clock frequencies of 1 GHz,  this approximates the number of cycles/event required for achieving pattern recognition at 40 MHz, and corresponds to a $\mathcal{O}(100)$ increase in processing efficiency over what attained by present or foreseen dedicated devices~\cite{SVT, FTK} and about a $\mathcal{O}(10^7)$ increase over unspecialized CPU-like architectures.\par
\begin{figure}[tbp] 
\centering
\includegraphics[width=.8\textwidth]{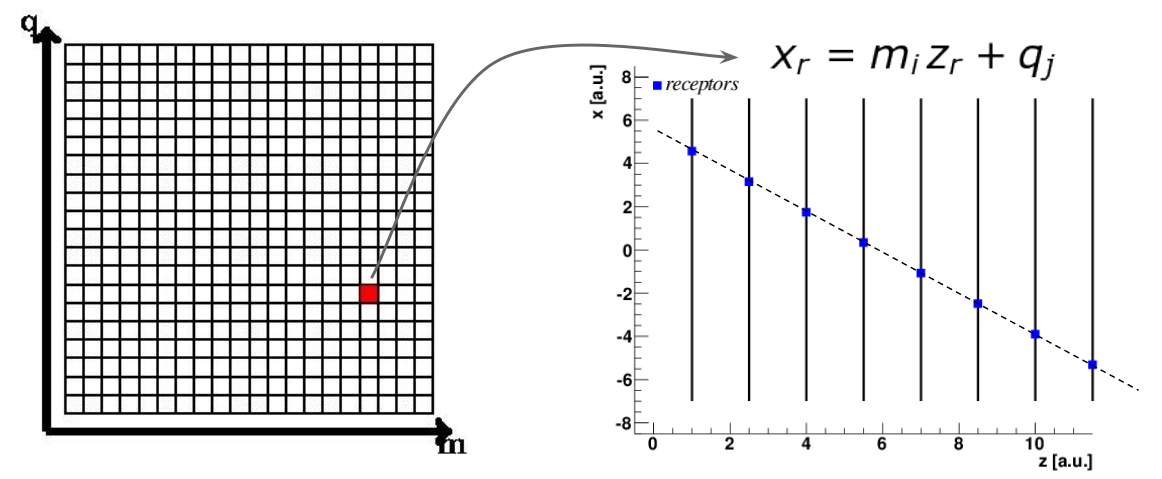}
\caption{Schematic representation of the detector mapping.
The parameter space (left panel) is discretized into cells; to each cell corresponds a track that intercepts the detector layers in a determined sequence of receptors (right panel).}
\label{fig:mapping2D}
\end{figure}
\begin{figure}[tbp]
\centering
\includegraphics[width=.4\textwidth]{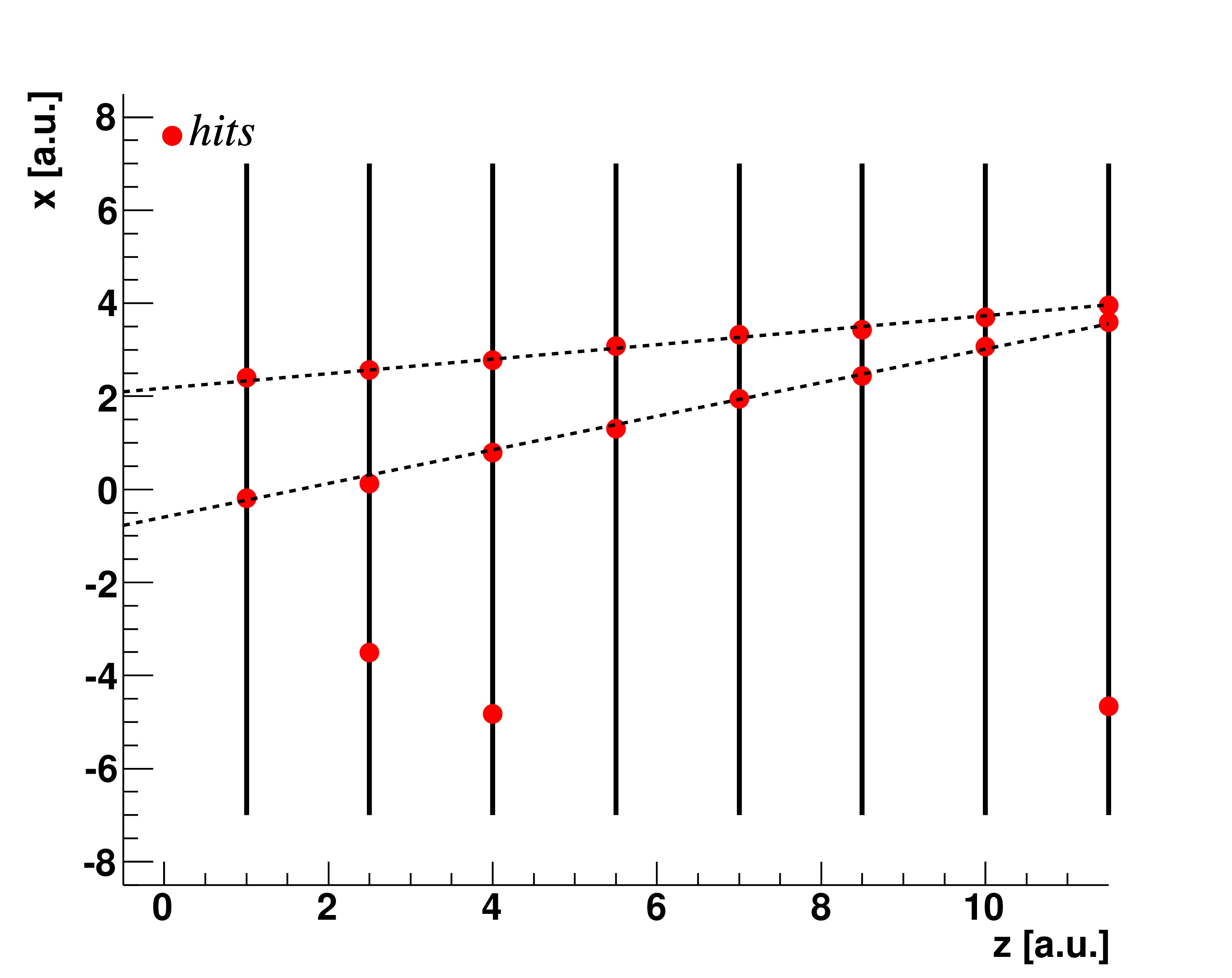}
\includegraphics[width=.4\textwidth]{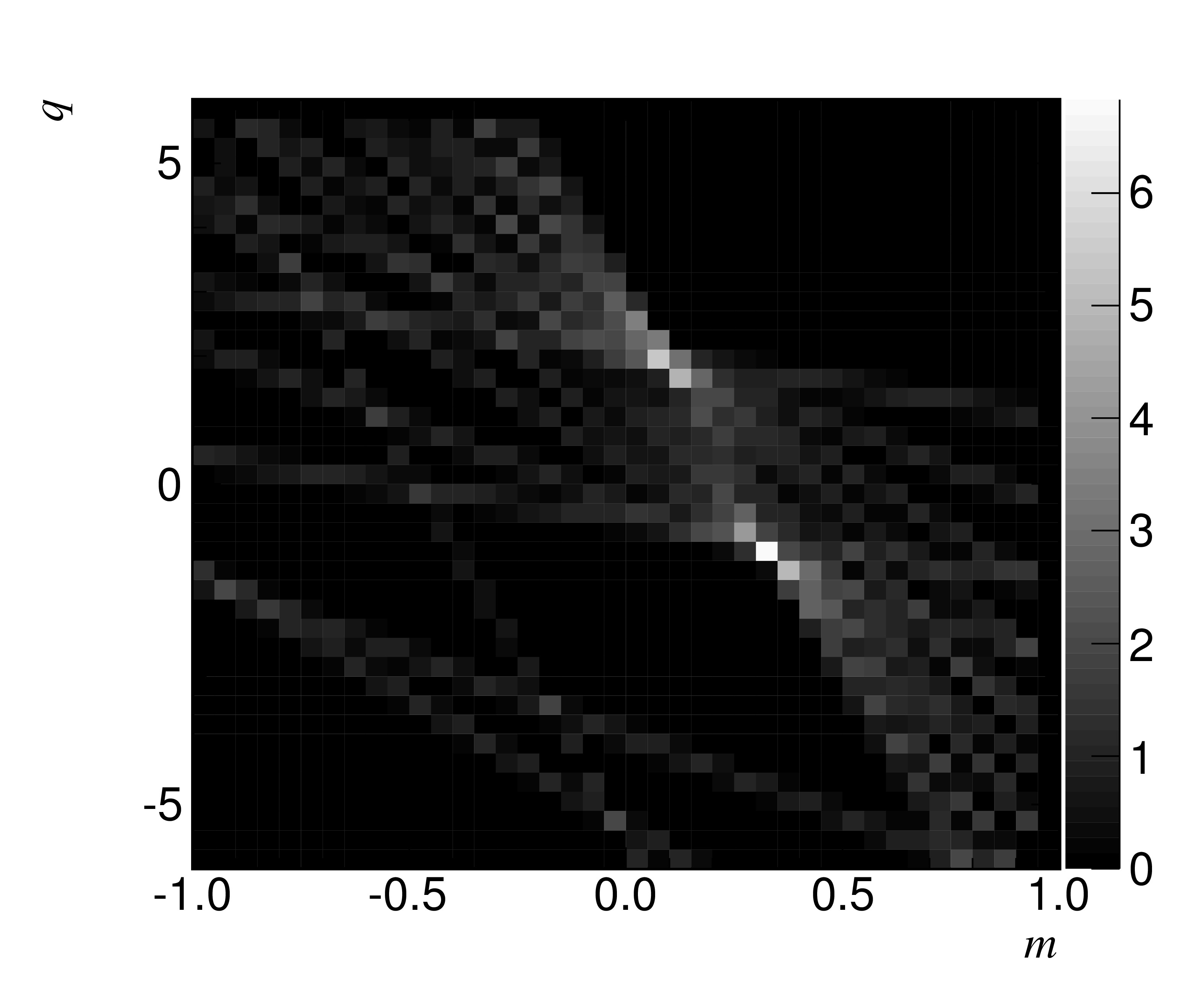}
\caption{Sketch of a simple event containing only two tracks and a few noise hits (left panel) and response of the retina (right panel).}
\label{fig:response_retina}
\end{figure}

The concept of a retina-inspired algorithm for track reconstruction is best understood using a simple example: a detector consisting of parallel layers of position-sensitive sensors that only measure one spatial coordinate, $x$. The trajectories of charged particles in the absence of magnetic field are straight lines identified by their angular coefficient $m$, and intercept $q$ with the $x$ axis in an arbitrary $(z,x)$ plane. We discretize the parameter space into \emph{cells} that mirror the visual receptive fields. The center of each cell identifies uniquely an ideal track in the detector space that intersects detector layers in spatial points called \emph{receptors}. Therefore the parameter-space cell $(m_i,q_j)$ maps into the set of receptors $\, x_k^{ij}$, where $k=1, \ldots, n$ runs over the detector layers (figure~\ref{fig:mapping2D}). This cell-receptors mapping is done for all cells of the track parameter space.
Once the receptors corresponding to all cells are known, the detector can be exposed to real tracks.  The distance $s_{\mathit{ijkr}} = \overline{x}_{k,r} -x_k^{\mathit{ij}}$ of the receptors from the observed hits is computed and the response of the $(m_i, q_j)$ retina-cell is calculated,
\begin{equation}
\label{eq:responsecell}
R_{\mathit{ij}} = \sum_{k,\,r} \exp\Big(-\frac{s^2_{\mathit{ijkr}}}{2\sigma^2}\Big),
\end{equation}
where $\overline{x}_{k,r}$  are the coordinates of the $r$th hit on the detector layer $k$, while 
$\sigma$ is a parameter of the algorithm. $R_{ij}$ represents the excitation of the receptive field.
The total response of the retina is obtained by calculating the excitation $R$ of all cells. Tracks are identified by local maxima among cells excited over a suitable threshold (figure~\ref{fig:response_retina}).\par 
In two dimensions, the algorithm bears analogies with the Hough transformation~\cite{Hough}. Generalization of the retina algorithm to the case of multiple dimensions, presence of magnetic field, and so forth is conceptually straightforward~\cite{PubNote}. After the track finding, determination of track parameters is refined using the excitation centroid of the nearest cells around each local maximum. This, along with the information contained in the smooth pattern-recognition response given by eq.~(\ref{eq:responsecell}), recovers resolution associated with the discretization of the parameter space and allows for coarse retina granularities with no penalty in performance. The total number of cells is mainly driven by the capability of separating similar tracks. The retina's continuous, analog-like response function and intrinsic capability of a fully parallel implementation down to hit level offer significant additional advantages. However, there is a significant complexity leap in passing from a proof of concept in a two-dimensional ideal case to a demonstration of feasibility in a real high-energy physics experiment with commercially available electronic components.

\section{Implementation}

\begin{figure}[tbp]
\centering
\includegraphics[width=.6\textwidth]{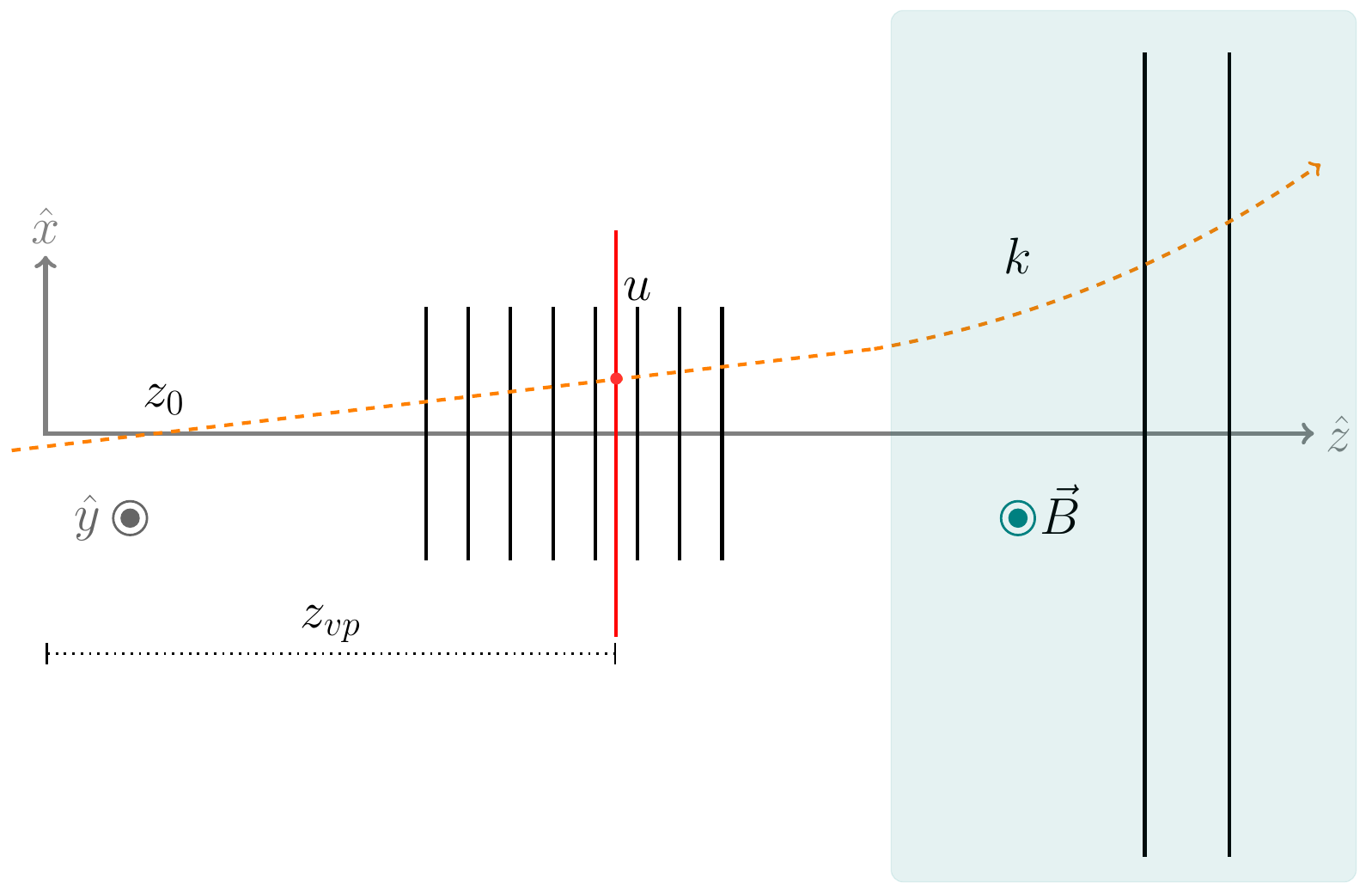}
\caption{Longitudinal view of the tracker volume layout in (not to scale). The dashed line indicates a track; the parallel black lines indicate planes of silicon detectors used in the retina algorithm; the red line indicates the primary ($u$, $v$) plane for track finding.}
\label{fig:layout}
\end{figure}

We implement the algorithm and simulate its performance in realistic conditions using the LHCb detector upgraded for the 2020 high-luminosity operations as a use case~\cite{PubNote}. The upgraded LHCb detector~\cite{FTDR} is a single-arm forward spectrometer covering polar angles from $0.8^{\degree}$  to $15.4^{\degree}$ from the beam. The detector is designed to study particles containing bottom or charm quarks produced in 14 TeV proton-proton collisions occurring every 25 ns at luminosities of 2--3$\times 10^{33}$ cm$^{-2}$s$^{-1}$, with 7.6--11.4 average interactions per crossing.  We use the standard LHCb upgrade simulation, which includes detailed modeling of a number of experimental effects, including multiple scattering, detector noise and so forth. The goal is to reconstruct, every 25 ns, 100--500 tracks, each associated to 10--20 hits that are scattered over 10--20 parallel detector planes. For this study, we choose an implementation of the retina that uses information from a small-angle telescope of ten tracking planes made of (i) eight layers of microvertex silicon detectors based on pixel technology~\cite{VELOTDR} and installed in a volume free of magnetic field and (ii) two additional layers of microstrip silicon detectors located upstream of the magnet~\cite{UTTDR} and immersed in its 0.05 T fringe field (Fig~\ref{fig:layout}).  Such telescope has about 50 mrad of angular acceptance and covers about half of the total tracking acceptance.  Complementing it with an additional, similarly sized telescope at large angles covers the full acceptance. More details on the configuration are in ref.~\cite{talk_Pietro}. Other configurations are possible, {\it e.g.}, including forward tracking detectors, depending on the target tracking performance needed. The logic is implemented in VHDL language; detailed logic-gate placement and simulation on the high-bandwidth Altera Stratix V device, model 5SGXEA7N2F45C2ES, is achieved using Altera's proprietary Quartus II software.\par
The realistic implementation of the algorithm poses two chief conceptual and technological challenges. One challenge is achieving an efficient distribution of the hit information coming from the detector to the processing engines that calculate the excitations. The need for a 40 MHz throughput with several Tbit/s of input data flow makes this a nontrivial task. 
The other challenge is performing pattern recognition quickly enough to remain within latency constraints. Solutions to either issues typically depend on the geometry of the tracking layout. We find an efficient solution based on the LHCb geometry, in which straight-line tracks traverse the vertex detector before being curved by the magnetic field and reach the downstream tracking stations. The idea is that tracking performance sufficient for triggering can be achieved by restricting the core of the pattern-recognition task to a region where the magnetic field is weak. This greatly simplifies the switching and the pattern recognition tasks.\par 
The switching is facilitated because regions of physically contiguous detector-hits tend to map into contiguous regions in track-parameter space, which have limited overlap. Hence, a mapping between detector hits and the parameters of possible tracks is obtained using simulation and results are used to associate a \emph{zip-code} with each possible detector hit. The zip-code is used by the nodes of the switching network to properly route each hit thus avoiding the need to feed all hits to all receptors. \par
\begin{figure}
\begin{center}
\includegraphics[trim=0cm 3.5cm 0cm 0cm, clip=true,scale= 0.5]{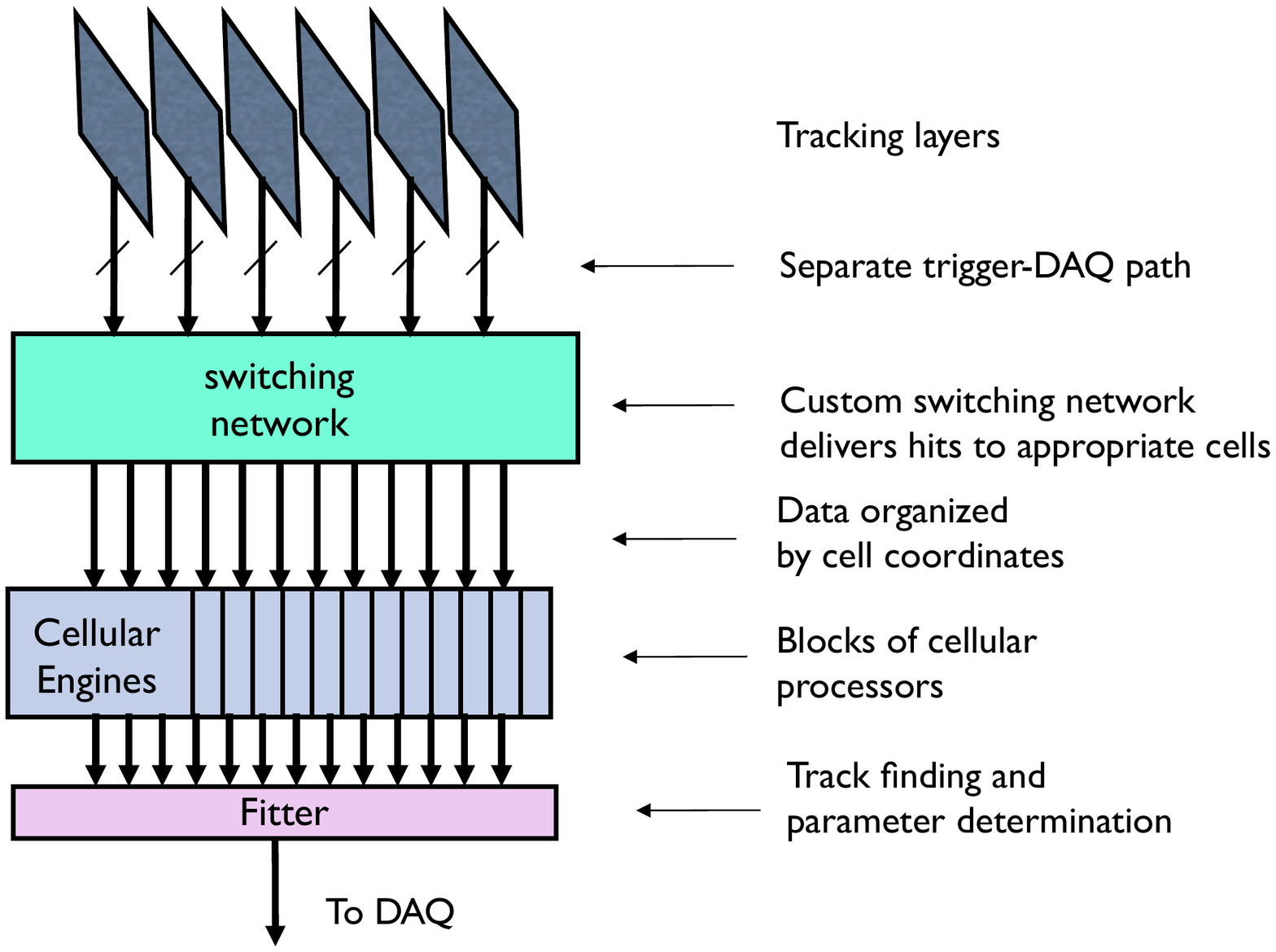}
\end{center}
\caption{Illustration of the device architecture.}
\label{fig:architecture}
\end{figure}

  The LHCb geometry allows factorizing the pattern recognition in full five-dimensional space of track parameters into two separate and simpler steps, independently of nonuniformities of the magnetic field or detector misalignments. First, tracks are assumed to be straight lines originated from a single nominal interaction point and track-finding is performed in a two-dimensional {\it primary} plane transverse to the beam, whose intersection which each track identifies the track's two primary cartesian parameters $u$ and $v$. Then, the determination of the momentum $p$ and the coordinates of the origin of the charged particle $d$ and $z$ are treated as small perturbations of the primary two-dimensional track. This is achieved by dividing the primary parameter space ($u$, $v$) in a fine grid of cells and, for each,  allow just for a small number of bins (lateral cells) in the three remaining track parameters. A track is first identified as a cluster over threshold in the primary plane and an estimate of all track parameters is then obtained by balancing the excitation found in the later cells for each compact dimension.
Figure~\ref{fig:architecture} illustrates the architecture of the retina track processor.

\subsection{Switching network}
\begin{figure}
\begin{center}
\includegraphics[width=0.45\textwidth]{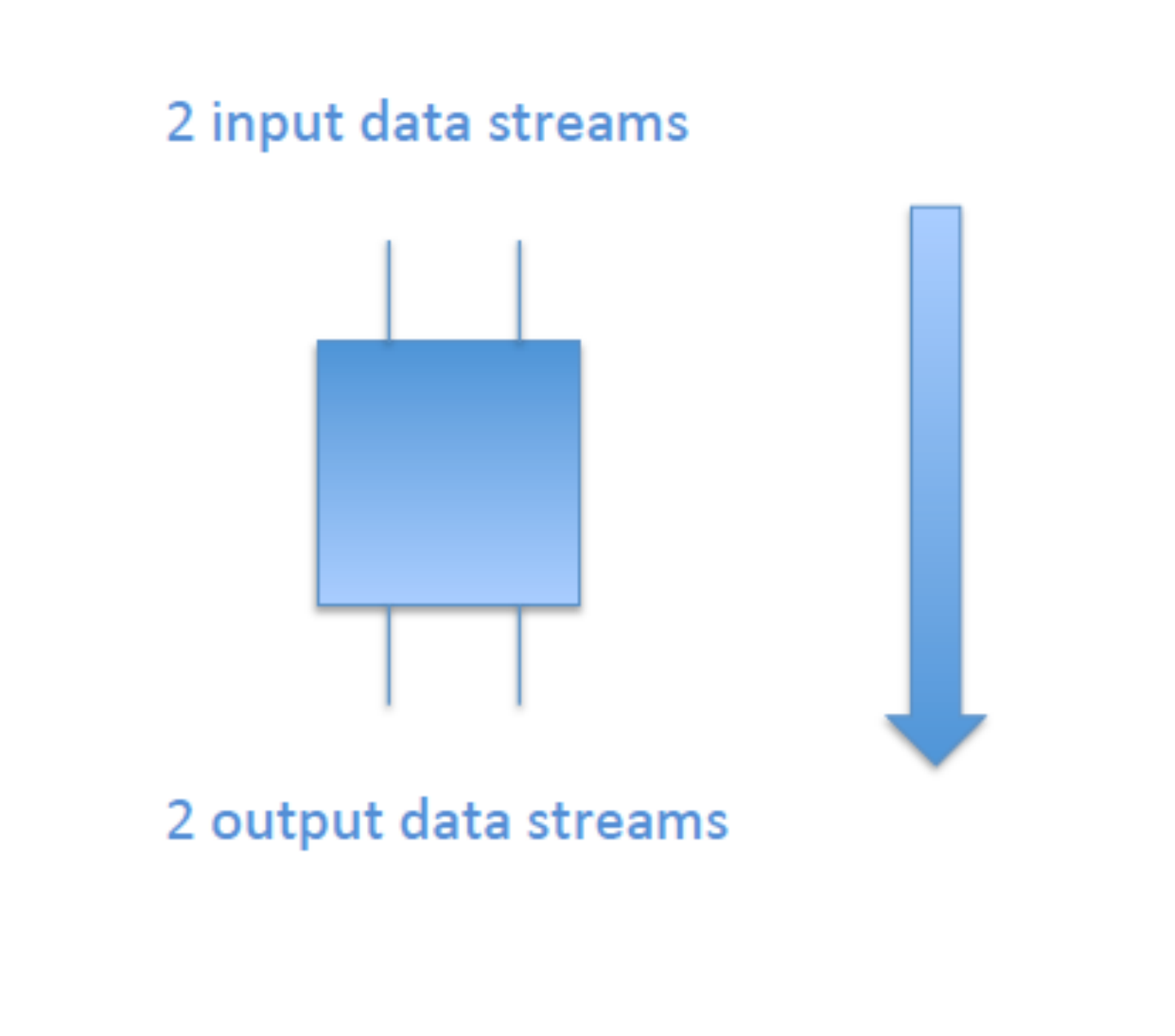}
\includegraphics[width=0.45\textwidth]{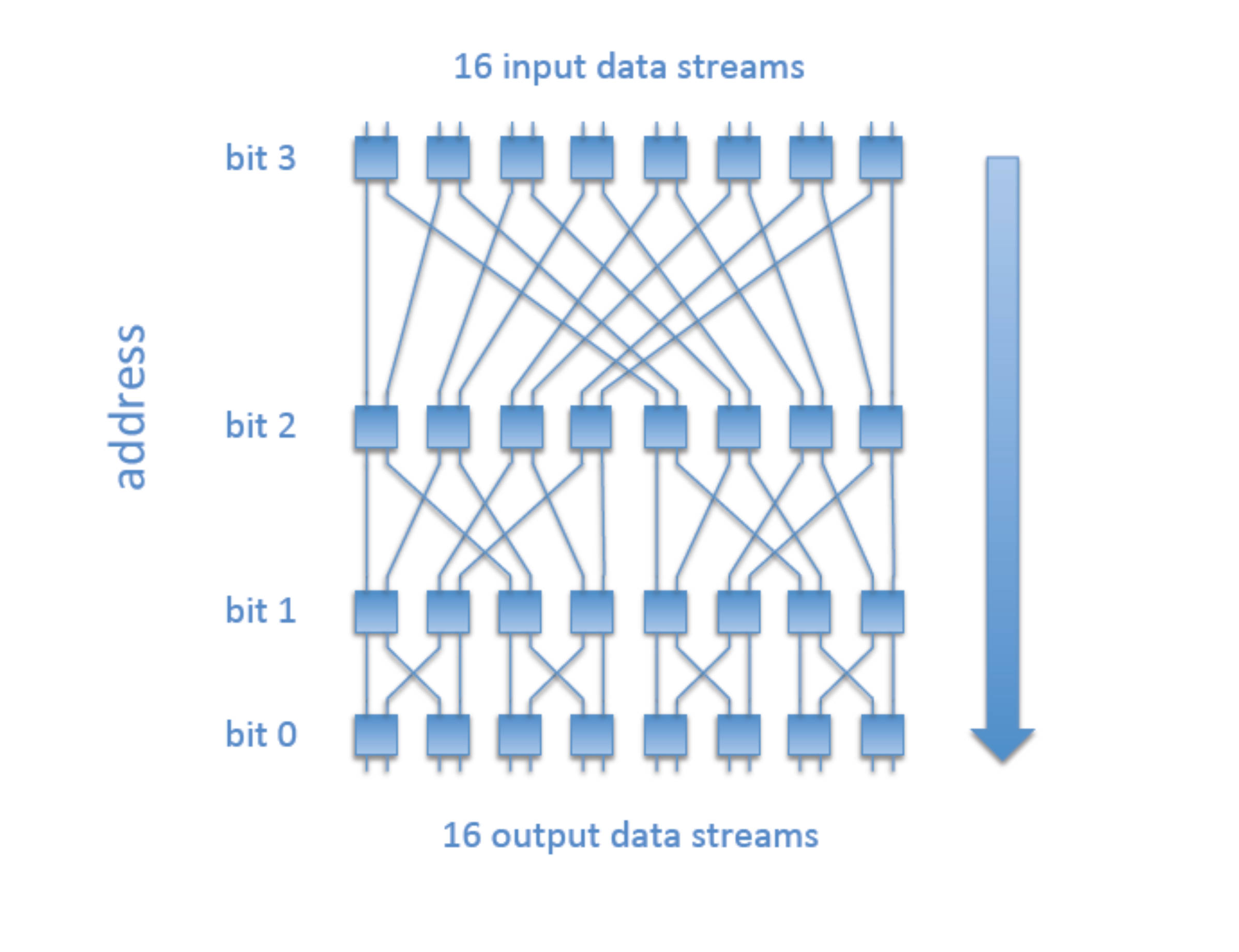}
\end{center}
\caption{Two-way sorter (left) and $16\times16$ switching network (right). Each of the network nodes are two-way sorters; the 4-bit address is referred to as zip-code in the text.} 
\label{fig:switch-concept}
\end{figure}

We design an intelligent and economical information-delivery system that routes each hit in parallel to all and only those engines for which such hit is likely to contribute a significant weight.  Each hit comes associated with a zip-code. At each stage of the switching network, each node reads the hit's zip-code and, based on a predetermined map loaded locally, routes the hit in parallel to the appropriate nodes (or engines) of the following stage. Each hit is dispatched and duplicated as prescribed by the map. For maximum efficiency, the switching logic is integrated in the same FPGA devices where the processing engines are hosted. 
The switch consists of a network of nodes, whose basic building blocks are two-way sorters (figure~\ref{fig:switch-concept}, left), with two input and two output data streams.
Left and right input data are merged and hits are dispatched to one or both outputs according to the zip-code carried by the incoming hit. If a stall from downstream layers occurs, one or both input streams are held.  Such elementary building blocks are combined to build the needed network topology, with the required switching capability (figure~\ref{fig:switch-concept}, right). A $N \times N$ network requires $N\log_2(N)/2$ elements. The modular structure allows scalability and reconfiguration of the system if necessary, and allows distributing the necessary addressing information over the whole network, storing the information only at the nodes where it is required.
 The switching network occupies approximately 10\% of the available logic in the Stratix V and completes its processing in 30 clock cycles. 
\subsection{The processing engine}

\begin{figure}
\begin{center}
\includegraphics[width=0.75\textwidth]{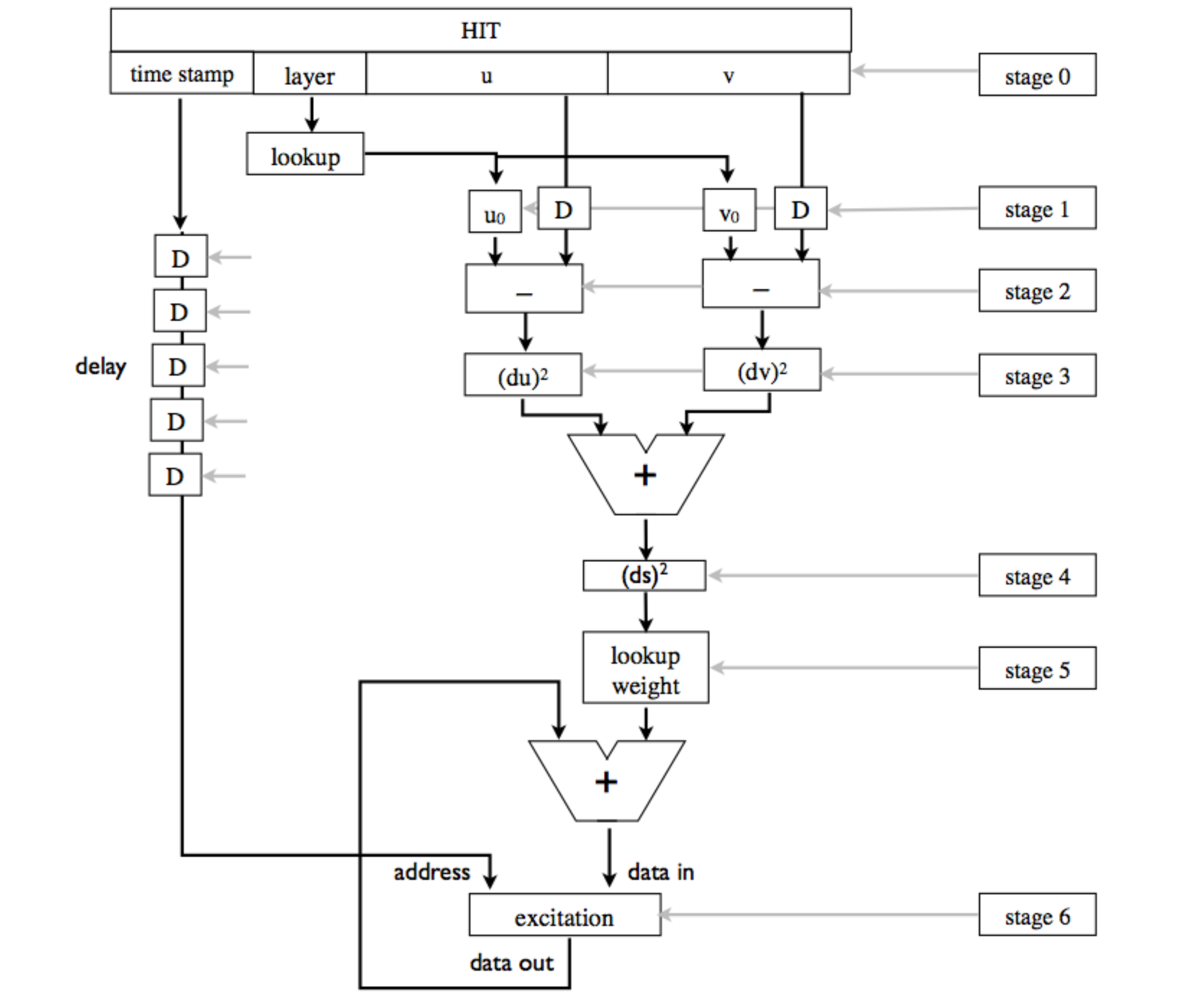}
\end{center}
\caption{Functional architecture of a receptor-field engine.}
\label{fig:engine}
\end{figure}

Each cell in parameter space is defined as a logic module, the engine. Each hit is defined as a 41 bits word encoding the hit's geometric coordinates, zip-code, and timestamp. 
The engine is implemented as a clocked pipeline (figure~\ref{fig:engine}).
The intersections $x_k$ and $y_k$ for each layer $k$ are stored in a read-only memory. The layer identifier associated with each incoming hit selects the appropriate set of $x_k$ and $y_k$ coordinates that are subtracted from the observed hit's $\bar{x}$ and $\bar{y}$ coordinates. The outcomes are squared and summed, and the result $R$ is rounded. A sigma function, common to all engines, is mapped into a ($8 \times 256$)-bit lookup table. The rounded $R$ is used as address to the lookup table. The outputs of the lookup table are accumulated for each hit of the event. The same hit is cycled seven times in the engine logic, once for the calculation of the excitation corresponding to the coordinates of the cell in the primary plane, and twice for each of the secondary track parameters, treated as perturbations. Hence, seven accumulators are defined for each cell. However, the excitation intensities contributed by each hit to all track parameters are computed in parallel, such that every engine is able to accept one hit every 20 ns approximately.  Several variants of the architecture are tested, from simple cases in which hits arrival is time-ordered to more complex scenarios where simultaneous processing of up to 16 events is allowed. Once readout of an event is completed, a word signaling the end of the event prompts each engine to share the content of its central cell to the neighboring engines. All engines in parallel compare the excitation in their central accumulator with the excitations received from the neighboring engines and raise a flag if they identify a  local maximum. Then, the coordinates and intensities of local maxima and the intensities of the nearest neighbors are output for track parameter extraction. \par In this scheme, each Stratix V can host up to 900 engines, leaving approximately 25\% of logic available for other uses, including 15\% of switching and the logic for center-of-excitation calculation.
This allows implementing a realistic retina-tracker based on a small-angle telescope with about 22~500 engines in about 32 chips, and a complete system of two-telescopes with about 50~000 engines in 64 chips. The maximum clock frequency is 350 MHz from a full Modelsim simulation.
\subsection{Clustering}
The logic that identifies the centers of excitation is a minor expansion. The clustering logic looks at the maximum flag and, if not busy, requires from the engine the content of all accumulators {\it and} the content of central accumulators of neighboring engines. These are used to calculate the track parameters as follows. The center-of-excitation calculation is factorized into two separate processes. The calculation restricted to the track parameters in the plane transverse to the beam line ($u,v$) implies finding the center of mass of a $3\times3$ square; the calculation relative to the remaining track parameters $(d,p,z)$ requires computing the center of mass of a $3\times 3\times 3$ cube. Only a subset of coordinates in each dimension is relevant for the final result; hence, the problem reduces to processing a smaller number of values. The operation for each coordinate is $u =  u_0/d_k + (\sum_{i,j} u_i l_{i,j}) / \sum l$, where $u_0/d_k$ is a global translation that depends on the absolute position of the engine and is not calculated in real time, but stored in a lookup table. Two distinct weights are simultaneously produced for $(u,v)$ and $(d,p,z)$, respectively. In a possible architecture, the computation of the center of excitation takes 11 clock cycles along with another 10 cycles for fanout with a logic that occupies a fraction not larger than 15\% of the Stratix V. To optimize resource sharing, a single center-of-excitation unit can serve multiple engines. Simulations show that a scheme with a unit serving each group of 12 engines is adequately sized for the hit occupancies expected. The search for the local maximum and the center of excitation use local copies of the accumulators so that the incoming hit flux is never stopped unless large time-fluctuations in the end event signal occur. In these cases the incoming hits are kept on hold and stored in the switch trees.

\subsection{Performance}
The achieved tracking performance is comparable with that of the offline reconstruction, as detailed in ref.~\cite{talk_Pietro}.
The full simulation shows that the device sustains an input frequency of 40 MHz of events, with the occupancy predicted by the full LHCb simulation, in the nominal luminosity conditions of the 2020 upgrade, $2\times10^{33}$ cm$^{-2}$s$^{-1}$. Contributions to latency are listed in table~\ref{tab:timing}. With the attained clock frequencies of up to 350 MHz, the latency for reconstructing online tracks is less than 0.5 $\mu$s, which is likely to be negligible compared with other latencies typically present in the DAQ. This makes the response of the device effectively {\it immediate}, thus making tracks available right after the tracking detectors have been read out. Tests simulating higher-track-density conditions show that the logic needed increases approximately linearly with the number of detector hits present in the event.

\begin{table}
\caption{Event-processing time in units of clock cycles of the Stratix V FPGA.}
\centering 
\begin{tabular}{lc}
\hline\hline
Task 					& 	Latency (cycles)	\\
\hline
Switching in readout board 	&	30				\\
Switching 	fanout		&	6				\\
Engine processing			&	70				\\
Clustering					&	11				\\
Output data				&	10				\\
\hline
Total						&	$< 150$			\\
\hline \hline
  \end{tabular}
  \label{tab:timing}
\end{table}


\begin{table}
\caption{Average retina occupancies and ghost-track reconstruction rates in two scenarios of instantaneous luminosities. As in Table I, `cluster' refers to local excitations in the retina, not silicon-detector clusters.}
\label{tab:hits_ghost}

\centering
\begin{tabular}{l c c}
\hline\hline
  & $2\times 10^{33}\rm cm^{-2}s^{-1}$ & $3\times 10^{33}\rm cm^{-2}s^{-1}$ \\
\hline
		Number of hits 				& 880 & 1220\\
		Number of clusters (over thrsh)	& 121 &  223\\
		Number of hits per engine 	& 1.3 &  1.95\\
		Ghost rate 				& 0.08 & 0.12 \\
\hline\hline
\end{tabular}
\end{table}

\section{Conclusions}
We report on the first realistic implementation of the artificial retina algorithm for track reconstruction. We simulate the full algorithm in high-end FPGA processors.
We determine the timing, occupancy, and ghost-rate performances of the algorithm using the conditions expected for the LHCb detector
upgraded for 2020 high-luminosity conditions, as a benchmark. The algorithm reconstruct tracks with offline-like efficiencies and resolutions~\cite{talk_Pietro} within a latency of less than 0.5~$\mu s$. This is 400 times faster than any existing or foreseen device used in HEP.  The observed latency is negligible with respect to other latencies present in the LHCb data acquisition system and makes the retina tracker to appear to DAQ as an additional detector that directly outputs tracks. This is the first demonstration of online reconstruction of tracks with offline-quality at 40 MHz~\cite{talk_Pietro, PubNote}.


\end{document}